\documentclass[12pt,a4paper]{article}

\usepackage[T1]{fontenc}
\usepackage[utf8]{inputenc}
\usepackage{amsmath}
\pdfoutput=1
\usepackage{amssymb}
\usepackage{amsfonts}

\usepackage{amsthm}
\usepackage{mathrsfs}
\usepackage{multicol}
\usepackage{makeidx}
\usepackage{graphicx}
\usepackage{xcolor}
\usepackage{float}

\usepackage{array}
\usepackage{pst-text}
\usepackage{pst-all}
\usepackage{pstricks-add}
\usepackage{pst-plot}
\usepackage{fancybox}
\usepackage{soul}
\usepackage{lineno}
\setstcolor{red}

\begin{document}
\title{Effective longitudinal wave through a random distribution of  poroelastic spheres in a poroelastic matrix  }
\author{Dossou GNADJRO, Amah S\'ena d'ALMEIDA }
\maketitle
\begin{center}
	 D\'epartement de Math\'ematiques, Universit\'e de Lom\'e, Togo\\	
\end{center}

	\begin{abstract}
	A random distribution of  poroelastic spheres  in a poroelastic medium obeying Biot's theory is considered. The scattering coefficients  of the fast and the slow waves are computed in the low frequency limit using the sealed pore boundary conditions. Analytical expressions of the\\ effective wavenumbers of the  coherent longitudinal waves (fast and slow ) are deduced in the Rayleigh limit using  a generalization of the Linton-Martin (LM) formula to poroelastic medium up to the order two in concentration.
	Some effective quantities (mass density, bulk modulus) of the heterogeneous media are estimated.
\end{abstract}

	\vspace{0.5cm}
	\noindent\textbf{Keywords:}  low frequency,  poroelastic medium, random medium, effective quantities

\section{Introduction}
The study of multiple scattering of acoustical waves by obstacles has\\ numerous applications in material sciences and in industry and considerable work has been done on the subject in the recent years. In scattering\\ problems, the significant parameter is $ka$ where $a$ is a characteristic length of the scatter and $k$ the wavenumber. The Rayleigh limit refers to cases where $ka\ll 1$. The medium supporting the wave can be a fluid, a solid or a porous medium. Many researchers have studied wave scattering by spherical inclusions suspended in a liquid \cite{1, 2, 6, 7, 14, 15, 19, 26}  or by spherical inclusions in an elastic or poroelastic medium \cite{4, 10, 16}. According to Biot's theory \cite{3} three kinds of waves, a fast, a slow  compressional waves and a transverse  one respectively denoted by (1), (2) and (t), can propagate in a porous medium. An encounter with an obstacle causes the scattering of any of them in the three kinds of waves by  mode conversion. The multiple scattering model of Waterman and Truell \cite{6}, which is an improvement of that of Foldy \cite{1} for spherical inclusions in a fluid, leads to an expression of the effective wavenumber for spheres randomly distributed in a fluid medium with a low density $ n_ {0} $ (the number of spheres per unit volume). This\\ formula is still valid in a porous medium for each kind of waves considered by neglecting the conversion phenomena \cite {10}. Foldy \cite{1},  Waterman and Truell formulas \cite{6} are derived from the Quasi Cristalline Approximation (QCA)  of Lax \cite{2}. Lloyd and Berry \cite{7} in 1967 extended the Waterman and Truell model to spherical inclusions of finite size by introducing the concept of \\radius of exclusion which prohibits two inclusions from interpenetration. This made it possible to correct the second order term in concentration by an\\ integral taking into account all the angles of scattering. In 2006, Linton and Martin \cite{14} proposed an equivalent writing of the effective wave number of LB by transforming the integral on the angles of scattering in double series. Later in 2012, Lupp\'e, Conoir and Norris \cite{16} extended the Linton and Martin model to solid or porous matrix taking into account the phenomenon of conversion. In \cite{22} the LCN model is used to study multiple scattering by a dilute distribution of elastic spheres in a poroelastic isotropic matrix. The same work is done here for a material formed of a poroelastic matrix in which are randomly distributed identical poroelastic spheres  of radius $a$ greater than the mean pore radius. \\ The paper is organized as follows. In  section 2 we  we state the  mathematical problem. In section 3 we calculate an analytical expression of the scattering coefficients in the Rayleigh limit using sealed pore conditions. Section 4  present    the  model of coherent wavenumbers of Lupp\'e, Conoir and Norris \cite{16} for poroelastic spheres obstacles in poroelastic media. In section 5,  we study the numerical results of scattering coefficients at low frequency. We also compare the velocities and attenuations of fast  and slow waves. In section 6, we derive the effective mass density and bulk modulus of the  coherent fast wave at the static limit using the model of Lupp\'e, Conoir and Norris (LCN).

\section{Statement of the mathematical problem}  
Consider a   poroelastic sphere of radius $a$ filled by a fluid and placed in a porous medium of infinite extension and saturated by another fluid. The mean pore radius of the matrix is denoted $a_{p}$ and $ a >> a_{p}$.   The sphere and the matrix have  different mechanical properties. The parameters  related to the sphere will be surmounted by a tilde in order to differentiate them from those of the matrix given in table 1.\\  The mass density of the porous medium is $ \rho = (1- \psi) \rho_{s} + \phi \rho_{0} $ ($ \rho_{s} $ mass density of the solid part,  $\rho_{0}$ mass density of the fluid and $ \phi $ the porosity) while the poroelastic sphere has mass density $\tilde{\rho}=(1-\tilde{\phi})\tilde{\rho}_{s}+\tilde{\phi} \tilde{\rho}_{0}$ ($\tilde{\rho}_{s}$  mass density of the solid part,  $\tilde{\rho}_{0}$ mass density of the  fluid   and $\tilde{\phi}$ porosity). A complete description of the porous medium according to Biot's theory includes  other parameters such as tortuosity and permeability \cite{3},\cite{11}.  The porous medium supports two longitudinal waves (fast and slow) of wavenumbers $k_{j}$, $j=1,2$ respectively and a transverse one of wavenumber $k_{t}$. When a plane harmonic wave of type $\alpha \in   \left\lbrace 1,2,t\right\rbrace  $ is incident on the poroelastic sphere, it gives rise to scattered spherical waves of type $\beta \in   \left\lbrace 1,2,t\right\rbrace  $ \cite{10}.\\ 
The physical space is related to the orthonormal reference $ (O, \vec{x}, \vec{y}, \vec{z}) $. The inclusion is modelled by a sphere of center $O$ and the incident plane wave is in the direction of $ \vec{z} $. Due to the geometry of the problem we express the spatial dependence of the harmonic potentials in the  spherical coordinates $r$, $\theta$, $\varphi$  associated to the local reference  $(O,\vec{e}_{r},\vec{e}_{\theta}, \vec{e}_{\varphi})$. We assume  $e^{-i\omega \tau}$ dependence in the time $\tau$ with $\omega$ the angular frequency. There is no dependence on $ \varphi $ when the fast or slow longitudinal wave is incident but the symmetry is lost when the transverse wave is incident  and the potentials depend on the variable $\varphi $\cite{16}. The scattering of  incident waves of wavenumbers $\displaystyle{k_{\alpha}=\frac{\omega}{c_{\alpha}}}$ ($\alpha=1,\;2,\;t$,  $c_{\alpha}$  phase velocity),  is described by scattering coefficients $ T_{n}^{\alpha \beta} $ for the mode $n$ where  $\beta= 1,\;2,\;t $ are  the three scattered waves generated  in the poroelastic matrix of wavenumbers $\tilde{k}_{\beta}=\frac{\omega}{\tilde{c}_{\beta}}$ \cite{16}. The far-field scattering function $f^{\alpha \beta}$ of a sphere is defined  in terms of $ T_{n}^{\alpha \beta} $ by 
\begin{equation}\label{e1}
f^{\alpha\beta}(\theta)=\sum_{n=0}^{+\infty}(2n+1)T_{n}^{\alpha \beta} P_{n}(\cos \theta) \; \text{ if } \alpha=1,\;2 
\end{equation}
for a longitudinal incident wave and
\begin{equation}\label{e2}
f^{\alpha\beta}(\theta)=\sum_{n=0}^{+\infty}\frac{2n+1}{n(n+1)}T_{n}^{\alpha \beta} P_{n}^{1}(\cos \theta)\cos \phi \; \text{ if } \alpha=t.
\end{equation}
for a transverse incident wave where $P_{n}$ is Legendre polynomial of order $n$ and $P_{n}^{1}$ is Legendre polynomial of order $n$ and 1 \cite{21}.
 Equation \eqref{e2}  shows that the scattering coefficient $T_{0}^{t \beta}$ $( \beta=1,\;2,\;t)$ must be ignored. The coefficient $T_{0}^{\alpha t}$ $(\alpha=1,\;2,\;t)$ must also be ignored due to the scattering  cross section argument by Yin and Truell \cite{4}.
\newpage
\begin{table}[H]
	\centering
	\begin{tabular}{|c|c|c|}
		\hline 
		Parameters   & QF20 \cite{19} & Stoll's sand \cite{20} \\ 
		\hline 
		\hline
		$K_{r}$ bulk modulus of solid grains (Pa)& $36,6.10^{9} $& $36,6.10^{9} $\\ 
		\hline 
		$K_{b}$ dried frame bulk modulus (Pa) & $9,47.10^{9}$ & $4,36.10^7$ \\ 
		\hline 
		$\mu$ dried frame shear modulus (Pa) & $7,63.10^{9}$ & $2,61.10^7$ \\ 
		\hline 
		$\rho_{s}$ mass density of solid grains ($kg.m^{-3}$) & $2760$ & $2650$\\ 
		\hline 
		$K_{0}$ bulk modulus of water (Pa) & $2,22.10^{9}$ & $2,22.10^{9}$\\ 
		\hline 
		$\rho_{0}$ mass density of water ($kg.m^{-3}$) & $1000$& $1000$ \\ 
		\hline 
		$\eta$ dynamic viscosity ($kg.m^{-1}.s^{-1}$)& $1,14.10^{-3}$& $1,14.10^{-3}$\\ 
		\hline 
		$\phi$ porosity & 0,402&0,47\\ 
		\hline 
		$\kappa$ permeability ($m^{2}$)& $1,68.10^{-11}$& $5.10^{-11}$ \\ 
		\hline 
		$a_{p}$ mean pore radius ($m$) & $3,26.10^{-5}$ & $10^{-5}$\\ 
		\hline 
		$\alpha$ tortuosity & 1,89 & 3  \\ 
		\hline 
	\end{tabular}
	\caption{\small{ Parameters of the fluid-saturated poroelastic materials}} 
\end{table}

 The phenomena of multiple scattering and mode conversion introduce differences in the behaviour of the displacement and the pressure fields in the poroelastic matrix and  sphere. In the poroelastic matrix , let $\sigma_{rr}$ and $\sigma_{r\theta}$ be the radial and tangential stresses, $u_{r}$ and $u_{\theta}$  the  radial and tangential displacement components  in  the solid frame, $w_{r}$ and $w_{\theta}$  the  radial and tangential  components of the relative fluid to solid displacement. Let also $\tilde{\sigma}_{rr}$ and $\tilde{\sigma}_{r\theta}$, $\tilde{u}_{r}$ and $\tilde{u}_{\theta}$ , $\tilde{w}_{r}$ and $\tilde{w}_{\theta}$ be the equivalent quantities in the poroelastic sphere.\\
\noindent  The application of the following sealed pores boundary conditions 
\begin{equation}\label{e3}
u_{r}= \tilde{u}_{r},\quad u_{\theta}= \tilde{u}_{\theta},\quad
w_{r}=0,\quad \tilde{w}_{r}=0,\quad \sigma _{rr}=\tilde{\sigma}_{rr},\quad 
\sigma_{r\theta }=\tilde{\sigma}_{r\theta }  
\end{equation} 
at the sphere interface $r = a$ for a mode $n$ allows to obtain a linear system of 6 equations to determine the 6 complex  unknowns  $T_{n}^{\alpha \beta}$ :
\begin{equation}\label{e4}
M_{n}. \vec{X}_{n}^{\alpha}=\vec{S}_{n}^{\alpha},
\end{equation}
The matrix $M_{n}$  depends on the spherical Hankel and Bessel functions of the first kind \cite{21},  $\alpha=1,\;2,\;t$ denotes  incident wave, $ \vec{X_{n}^{\alpha}}$ is an unknown vector containing both the scattering coefficients  $T_{n}^{\alpha \beta}$ ( $\alpha,\,\beta=1,\,2,\,t$) and the modal coefficients generated in the inclusion $\tilde{T}_{n}^{\alpha \beta}$ ( $\alpha,\,\beta=1,\,2,\,t$); $\vec{S_{n}^{\alpha}}$  is the source column vector (see appendix \label{ap1}).

\section{Determination of scattering coefficients}
Let $x_{\alpha}=k_{\alpha}a$ and $\tilde{x}_{\alpha}=\tilde{k}_{\alpha}a$ ($\alpha=1,\,2,\,t$) be  dimensionless wavenumbers. At low frequency limit ($\left|x_{\alpha}\right|<<1$, $\left|\tilde{x}_{\alpha}\right|<<1$),  using Taylor series development and extracting the dominant terms, the computations of the scattering\\ coefficients show that $T_{0}^{\alpha \beta}$ ($\alpha,\; \beta=1,\,2$),  $T_{1}^{\alpha \beta}$, $T_{2}^{\alpha \beta}$ $(\alpha,\;\beta=1,\;2,\; t)$ are $O(x^{m}\tilde{x}^{p})$ and $T_{n}^{\alpha \beta}$ $(\alpha,\;\beta=1,\;2,\; t)$ is $O(x^{5})$ for $n\geq 3$, 
 where   $m$ and $p$ are integers such that $m+p=3$ and the notation $x^{5}$ means the product $x_{1}^{q_{1}}x_{2}^{q_{2}}x_{t}^{q_{3}}\tilde{x}_{t}^{q_{4}}\tilde{x}_{2}^{q_{5}}\tilde{x}_{t}^{q_{6}}$ where $q_{1}$, $q_{2}$, $q_{3}$, $q_{4}$, $q_{5}$ and $q_{6}$  are integers such that\\ $q_{1}+q_{2}+q_{3}+q_{4}+q_{5}+q_{6}=5$.\\ 
We introduce  the following quantities:
 \begin{equation}\label{e301}
 \gamma_{\alpha\beta}=\gamma_{\alpha}-\gamma_{\beta},\quad \tilde{\gamma}_{\alpha \beta}= \tilde{\gamma}_{\alpha}-\tilde{\gamma}_{\beta}\quad (\alpha,\,\beta=1,\,2,\,t)
 \end{equation}
 	\begin{equation}\label{e401}
 	\mathcal{H}_{\alpha}=H-2\mu+\gamma_{\alpha}C,\quad 
 	\tilde{\mathcal{H}}_{\alpha}=\tilde{H}-2\tilde{\mu}+\tilde{\gamma}_{\alpha}\tilde{C},\quad (\alpha=1,\,2)
 	\end{equation}
 \begin{equation}\label{e5}
 a_{\alpha}=(\mathcal{H}_{\alpha}+2\mu)(\mu-\tilde{\mu}),\; 
 b_{\alpha}=(\mathcal{H}_{\alpha}+2\mu)(3\tilde{\mu}+2\mu)\; (\alpha=1,\,2)
 \end{equation}
 \begin{equation}
  c_{\alpha}=\mu(5\tilde{\mathcal{H}}_{\alpha}+4\mu+6\tilde{\mu}) \; (\alpha=1,\,2)
 \end{equation}
 \begin{equation}\label{e6}
 e_{\alpha}=(\mu-\tilde{\mu})(12\mathcal{H}_{\alpha}-7\tilde{\mathcal{H}}_{\alpha}+16\mu-6\tilde{\mu}) ,\quad (\alpha=1,\,2)
 \end{equation}
 \begin{equation}\label{e7}
 f_{\alpha\beta}=(\mu-\tilde{\mu})(12\mathcal{H}_{\alpha}-7\tilde{\mathcal{H}}_{\beta}+16\mu-6\tilde{\mu}) ,\quad (\alpha\neq \beta=1,\,2)
 \end{equation}
 \begin{equation}\label{e8}
 g_{\alpha}=7\tilde{\mathcal{H}}_{\alpha}(3\mu+2\tilde{\mu})+8\mu^2+50\mu \tilde{\mu}+12\tilde{\mu}^2,\quad (\alpha=1,\,2)
 \end{equation}
 \begin{equation}\label{e9}
 h_{\alpha}=12\mathcal{H}_{\alpha}(2\mu+3\tilde{\mu})+32\mu^2+75\mu \tilde{\mu}-2\tilde{\mu}^2,\quad (\alpha=1,\,2)
 \end{equation}
 \begin{equation}\label{e10}
r_{\alpha}=\tilde{\mathcal{H}}_{\alpha}(2\mu+3\tilde{\mu})+2\tilde{\mu}(4\mu+\tilde{\mu}),\quad (\alpha=1,\,2)
 \end{equation}
 \begin{equation}\label{e11}
 	s_{\alpha}=7\tilde{\mathcal{H}}_{\alpha}(\mu-\tilde{\mu})+8\mu^2-2\mu \tilde{\mu}-6\tilde{\mu}^2, \quad (\alpha=1,\,2)
 \end{equation}
 \begin{equation}\label{e12}
 	d=\mu(\mu-\tilde{\mu}),\; l=(\mu-\tilde{\mu})(4\mu+\tilde{\mu}),\; z=(\mu-\tilde{\mu})(\tilde{\mathcal{H}}_{1}-\tilde{\mathcal{H}}_{2}),\; w=(\mu-\tilde{\mu})(8\mu-\tilde{\mu})
 \end{equation}
 \noindent Above ($H$, $C$, $\mu$) and ($\tilde{H}$, $\tilde{C}$, $\tilde{\mu}$) are poroelastic moduli  characterizing the behavior of the matrix and of the sphere  respectively \cite{9} \cite{19}, $\gamma_{\alpha}$ and $\tilde{\gamma}_{\alpha}$ are\\ compatibility coefficients \cite{16}. The analytic forms of the scattering\\ coefficients are listed below :\\
	$\bullet$ For the mode $n=0$ :
	\begin{equation}\label{e13}
	T_{0}^{11} =\frac{i x_{1}^{3}}{3}  \left(B_{0}^{11} -1\right),\quad T_{0}^{12} =\frac{i x_{1}^{2} x_{2}}{3}  B_{0}^{12} 
	\end{equation}
	\begin{equation}\label{e14}
	T_{0}^{21} =\frac{i x_{1} x_{2}^{2} }{3} B_{0}^{21},\quad 	T_{0}^{22} =\frac{i x_{2}^{3}}{3}  \left(B_{0}^{22} -1\right)
	\end{equation}
		where 
		\begin{equation}\label{e15}
		B_{0}^{11} =\frac{3\gamma_{2}\tilde{\gamma}_{12}\left(\mathcal{H}_{1}+2\mu\right)}{\Delta_{0}},\quad 	B_{0}^{12} = \frac{3\gamma_{1}\tilde{\gamma}_{21}\left(\mathcal{H}_{1}+2\mu\right)}{\Delta_{0}}
		\end{equation}		
			\begin{equation}\label{e16}
				B_{0}^{21} =\frac{3\gamma_{2}\tilde{\gamma}_{12}\left(\mathcal{H}_{2}+2\mu\right)}{\Delta_{0}}, \quad 	B_{0}^{22} = \frac{3\gamma_{1}\tilde{\gamma}_{21}\left(\mathcal{H}_{2}+2\mu\right)}{\Delta_{0}}
			\end{equation}
				\begin{equation}\label{e17}
				\Delta_{0}=\gamma_{21}\left[\tilde{\gamma}_{1}\left(3\mathcal{\tilde{H}}_{2}+4\mu+2\tilde{\mu}\right)-\tilde{\gamma}_{2}\left(3\mathcal{\tilde{H}}_{1}+4\mu+2\tilde{\mu}\right)\right]
				\end{equation}	
		$\bullet$ For the mode $n=1$ :			
	\begin{equation}\label{e18}
	T_{1}^{11} =-\frac{i x_{1}^{3}}{3}  B_{1}^{11},\quad  T_{1}^{12} =-\frac{i x_{1}x_{2}^{2}}{3}  B_{1}^{12},\quad T_{1}^{1t} =-\frac{i x_{1}x_{t}^{2}}{3}  B_{1}^{1t}
	\end{equation}  	
	\begin{equation}\label{e19}
	T_{1}^{21} =-\frac{i  x_{2}x_{1}^{2}}{3}  B_{1}^{21},\quad  T_{1}^{22} =-\frac{i x_{2}^{3}}{3}  B_{1}^{22},\quad T_{1}^{2t} =-\frac{i x_{2}x_{t}^{2}}{3}  B_{1}^{2t}
	\end{equation}	
	\begin{equation}\label{e20}
		T_{1}^{t1} =-\frac{i  x_{t}x_{1}^{2}}{3}  B_{1}^{t1},\quad  T_{1}^{t2} =-\frac{i  x_{t}x_{2}^{2}}{3}  B_{1}^{t2},\quad T_{1}^{tt} =-\frac{i x_{t}^{3}}{3}  B_{1}^{tt}
	\end{equation}
		where
	\begin{equation}\label{e21}
	\begin{array}{lcl}
	B_{1}^{11} &= &\left[4\left(\tilde{\gamma}_{t2}+\tilde{\gamma}_{1t}\frac{\tilde{x}_{2}^{2}}{\tilde{x}_{1}^{2}}\right)\left(2\gamma_{2t}a_{1}\frac{x_{1}^{2}}{x_{t}^{2}}+\gamma_{1}a_{2}\frac{x_{2}^{2}}{x_{t}^{2}}\right)+\tilde{\gamma}_{12}\left(\gamma_{1}b_{2}\frac{x_{2}^{2}}{x_{t}^{2}}+2\gamma_{2t}b_{1}\frac{x_{1}^{2}}{x_{t}^{2}}\right)\frac{\tilde{x}_{t}^{2}}{\tilde{x}_{1}^{2}}\right.\\ & & \left.+2\gamma_{1}\left(\tilde{\gamma}_{t2}c_{1}+\tilde{\gamma}_{1t}c_{2}\frac{\tilde{x}_{2}^{2}}{\tilde{x}_{1}^{2}}\right)\ +2\gamma_{2t}\left(\tilde{\gamma}_{2}r_{1}-\tilde{\gamma}_{1}r_{2}\frac{\tilde{x}_{2}^{2}}{\tilde{x}_{1}^{2}}\right)\frac{\tilde{x}_{t}^{2}}{x_{t}^{2}}\right.\\  & &\left.+4\gamma_{1}\tilde{\gamma}_{12}d \frac{\tilde{x}_{t}^{2}}{\tilde{x}_{1}^{2}}+8\tilde{\gamma}_{t}\gamma_{2t}z\frac{\tilde{x}_{2}^{2}}{x_{t}^{2}}\right] / \Delta_{1}
	\end{array}
	\end{equation}
	\begin{equation}\label{e22}
	\begin{array}{lcl}
		B_{1}^{12}& = & \left[4(3\gamma_{1}-2\gamma_{s})a_{1} \left(\tilde{\gamma}_{2t}+\tilde{\gamma}_{t1}\frac{\tilde{x}_{2}^2}{\tilde{x}_{1}^{2}}\right)\frac{x_{1}^2}{x_{t}^{2}}+\tilde{\gamma}_{12}b_{1}\left(2\gamma_{t}-3\gamma_{1}\right)\frac{\tilde{x}_{t}^2}{\tilde{x}_{1}^{2}}\frac{x_{1}^2}{x_{t}^{2}}\right.\\  & &\left.+2\gamma_{1}\left(\tilde{\gamma}_{2t}c_{1}+\tilde{\gamma}_{t1}c_{2}\frac{\tilde{x}_{2}^2}{\tilde{x}_{1}^2}\right)+4\gamma_{1}\tilde{\gamma}_{21}d\frac{\tilde{x}_{t}^2}{\tilde{x}_{1}^2}+2\tilde{\gamma}_{2}\gamma_{t1}r_{1}\frac{\tilde{x}_{t}^2}{x_{t}^2}\right.\\ & &\left.+2\tilde{\gamma}_{1}\gamma_{1t}r_{2}\frac{\tilde{x}_{t}^2}{x_{t}^2}\frac{\tilde{x}_{2}^2}{\tilde{x}_{1}^2}+8\tilde{\gamma}_{t}\gamma_{1t}z\frac{\tilde{x}_{2}^2}{x_{t}^2}\right]/\Delta_{1}
	\end{array}
	\end{equation}
	\begin{equation}\label{e23}
	\begin{array}{lcl}
	B_{1}^{1t} &= &\left[4\left(\tilde{\gamma}_{t2}+\tilde{\gamma}_{1t}\frac{\tilde{x}_{2}^{2}}{\tilde{x}_{1}^{2}}\right)\left(\gamma_{1} a_{2}\frac{x_{2}^{2}}{x_{t}^{2}}+(2\gamma_{2}-3\gamma_{1})a_{1}\frac{x_{1}^{2}}{x_{t}^{2}}\right)\right.\\  & &\left.+2\gamma_{21}\left(\tilde{\gamma}_{2}r_{1}-\tilde{\gamma}_{1}r_{2}\frac{\tilde{x}_{2}^{2}}{\tilde{x}_{1}^{2}}\right)\frac{\tilde{x}_{t}^{2}}{x_{t}^{2}} +8\tilde{\gamma}_{t}\gamma_{21}z\frac{\tilde{x}_{2}^{2}}{x_{t}^{2}}\right.\\ & & \left. +\tilde{\gamma}_{12}\left(\gamma_{1}b_{2}\frac{x_{2}^{2}}{x_{t}^{2}}+(2\gamma_{2}-3\gamma_{1})b_{1}\frac{x_{1}^{2}}{x_{t}^{2}}\right)\frac{\tilde{x}_{t}^{2}}{\tilde{x}_{1}^{2}}\right] / \Delta_{1}
	\end{array}
	\end{equation}
	\begin{equation}\label{e24}
	\begin{array}{lcl}
	B_{1}^{21}& = & \left[4 (3\gamma_{2}-2\gamma_{t})a_{2}\left(\tilde{\gamma}_{t2}+\tilde{\gamma}_{1t}\frac{\tilde{x}_{2}^2}{\tilde{x}_{1}^{2}}\right)\frac{x_{2}^2}{x_{t}^{2}}+\tilde{\gamma}_{21}\left(2\gamma_{t}-3\gamma_{2}\right)b_{2}\frac{\tilde{x}_{t}^2}{\tilde{x}_{1}^{2}}\frac{x_{2}^2}{x_{t}^{2}}\right.\\ & &\left.+4\gamma_{2}\tilde{\gamma}_{12}d\frac{\tilde{x}_{t}^2}{\tilde{x}_{1}^2}+2\tilde{\gamma}_{2}\gamma_{t2}r_{1}\frac{\tilde{x}_{t}^2}{x_{t}^2}+2\tilde{\gamma}_{1}\gamma_{2t}\frac{\tilde{x}_{t}^2}{x_{t}^2}\frac{\tilde{x}_{2}^2}{\tilde{x}_{1}^2}r_{2}+8\tilde{\gamma}_{t}\gamma_{t2}z\frac{\tilde{x}_{2}^2}{x_{t}^2}\right.\\ & & \left.+2\gamma_{2}\left(\tilde{\gamma}_{t2}c_{1}+\frac{\tilde{x}_{2}^2}{\tilde{x}_{1}^2}\tilde{\gamma}_{1t}c_{2}\right) \right]/\Delta_{1}
	\end{array}
	\end{equation}
	\begin{equation}\label{e25}
		\begin{array}{lcl}
		B_{1}^{22} &= &\left[4\left(\tilde{\gamma}_{2t}+\tilde{\gamma}_{t1}\frac{\tilde{x}_{2}^{2}}{\tilde{x}_{1}^{2}}\right)\left(2\gamma_{1t}a_{2}\frac{x_{1}^{2}}{x_{t}^{2}}+\gamma_{2}a_{1}\frac{x_{1}^{2}}{x_{t}^{2}}\right)\right.\\  & &\left. +\tilde{\gamma}_{21}\left(2\gamma_{1t}b_{2}\frac{x_{2}^{2}}{x_{t}^{2}}+\gamma_{2}b_{1}\frac{x_{1}^{2}}{x_{t}^{2}}\right)\frac{\tilde{x}_{t}^{2}}{\tilde{x}_{1}^{2}}+2\gamma_{2}\left(\tilde{\gamma}_{2t}c_{1}+\tilde{\gamma}_{t1}c_{2}\frac{\tilde{x}_{2}^{2}}{\tilde{x}_{1}^{2}}\right) \right.\\ & & \left.+4\gamma_{2}\tilde{\gamma}_{21}d \frac{\tilde{x}_{t}^{2}}{\tilde{x}_{1}^{2}}+2\gamma_{t1}\left(\tilde{\gamma}_{2}r_{1}-\tilde{\gamma}_{1}r_{2}\frac{\tilde{x}_{2}^{2}}{\tilde{x}_{1}^{2}}\right)\frac{\tilde{x}_{t}^{2}}{x_{t}^{2}}+8\tilde{\gamma}_{t}\gamma_{t1}z\frac{\tilde{x}_{2}^{2}}{x_{t}^{2}}\right] / \Delta_{1}
		\end{array}
	\end{equation}
	\begin{equation}\label{e26}
		\begin{array}{lcl}
		B_{1}^{2t} &= &\left[4\left(\tilde{\gamma}_{2t}+\tilde{\gamma}_{t1}\frac{\tilde{x}_{2}^{2}}{\tilde{x}_{1}^{2}}\right)\left(\gamma_{2} a_{1}\frac{x_{1}^{2}}{x_{t}^{2}}+(2\gamma_{1}-3\gamma_{2})a_{2}\frac{x_{2}^{2}}{x_{t}^{2}}\right)\right.\\  & &\left. +\tilde{\gamma}_{21}\left(\gamma_{2}b_{1}\frac{x_{1}^{2}}{x_{t}^{2}}+(2\gamma_{1}-3\gamma_{2})b_{2}\frac{x_{2}^{2}}{x_{t}^{2}}\right)\frac{\tilde{x}_{t}^{2}}{\tilde{x}_{1}^{2}} \right.\\ & & \left.+2\gamma_{21}\left(\tilde{\gamma}_{2}r_{1}-\tilde{\gamma}_{1}r_{2}\frac{\tilde{x}_{2}^{2}}{\tilde{x}_{1}^{2}}\right)\frac{\tilde{x}_{t}^{2}}{x_{t}^{2}}+8\tilde{\gamma}_{t}\gamma_{21}z\frac{\tilde{x}_{2}^{2}}{x_{t}^{2}}\right] / \Delta_{1}
		\end{array}
	\end{equation}
	\begin{equation}\label{e27}
	\begin{array}{lcl}
	B_{1}^{t1} &= &\left[ 8\gamma_{t}a_{2}\left(\tilde{\gamma}_{t2}+\tilde{\gamma}_{1t}\frac{\tilde{x}_{2}^2}{\tilde{x}_{1}^2}\right)\frac{x_{2}^2}{x_{t}^2}+4\gamma_{2}\left(\tilde{\gamma}_{t2}c_{1}+\tilde{\gamma}_{1t}c_{2}\frac{\tilde{x}_{2}^2}{\tilde{x}_{1}^2}\right) \right.\\& &\left. +\tilde{\gamma}_{12}\gamma_{t}b_{2}\frac{\tilde{x}_{t}^2}{\tilde{x}_{1}^2}\frac{x_{2}^2}{x_{t}^2}+8\tilde{\gamma}_{12}\gamma_{2}d\frac{\tilde{x}_{t}^2}{\tilde{x}_{1}^2} \right.\\ & & \left. +4\gamma_{2t}\left(\tilde{\gamma}_{2}r_{1}-\tilde{\gamma}_{1}r_{2}\frac{\tilde{x}_{2}^{2}}{\tilde{x}_{1}^{2}}\right)\frac{\tilde{x}_{t}^{2}}{x_{t}^{2}}+16\tilde{\gamma}_{t}\gamma_{2t}z\frac{\tilde{x}_{2}^{2}}{x_{t}^{2}}\right]/\Delta_{1}
	\end{array}
	\end{equation}
	\begin{equation}\label{e28}
	\begin{array}{lcl}
	B_{1}^{t2} &= &\left[ 8\gamma_{t}a_{1}\left(\tilde{\gamma}_{2t}+\tilde{\gamma}_{t1}\frac{\tilde{x}_{2}^2}{\tilde{x}_{1}^2}\right)\frac{x_{2}^2}{x_{t}^2}+4\gamma_{1}\left(\tilde{\gamma}_{2t}c_{1}+\tilde{\gamma}_{t1}c_{2}\frac{\tilde{x}_{2}^2}{\tilde{x}_{1}^2}\right) \right.\\& &\left. +\tilde{\gamma}_{21}\gamma_{t}b_{1}\frac{\tilde{x}_{t}^2}{\tilde{x}_{1}^2}\frac{x_{1}^2}{x_{t}^2}+8\tilde{\gamma}_{21}\gamma_{1}d\frac{\tilde{x}_{t}^2}{\tilde{x}_{1}^2}\right.\\ & & \left. +4\gamma_{t1}\left(\tilde{\gamma}_{2}r_{1}-\tilde{\gamma}_{1}r_{2}\frac{\tilde{x}_{2}^{2}}{\tilde{x}_{1}^{2}}\right)\frac{\tilde{x}_{t}^{2}}{x_{t}^{2}}+16\tilde{\gamma}_{t}\gamma_{t1}z\frac{\tilde{x}_{2}^{2}}{x_{t}^{2}}\right]/\Delta_{1}
	\end{array}
	\end{equation}
	\begin{equation}\label{e29}
		\begin{array}{lcl}
		B_{1}^{tt} &= &\left[8\gamma_{s}\left(\tilde{\gamma}_{t2}+\tilde{\gamma}_{1t}\frac{\tilde{x}_{2}^{2}}{\tilde{x}_{1}^{2}}\right)\left(a_{2}-a_{1}\frac{x_{1}^{2}}{x_{t}^{2}}\right)\frac{x_{2}^{2}}{x_{t}^{2}}\right.\\  & &\left. +2\gamma_{t}\tilde{\gamma}_{12}\left(b_{2}-b_{1}\frac{x_{1}^{2}}{x_{t}^{2}}\right)\frac{\tilde{x}_{2}^{2}}{\tilde{x}_{t}^{2}}\frac{\tilde{x}_{t}^2}{\tilde{x}_{1}^2}+4\gamma_{12}\left(\tilde{\gamma}_{2t}c_{1}-\tilde{\gamma}_{1t}c_{2}\frac{\tilde{x}_{2}^{2}}{\tilde{x}_{1}^{2}}\right)\right.\\ & & \left. +8\gamma_{12}\tilde{\gamma}_{21}d \frac{\tilde{x}_{t}^{2}}{\tilde{x}_{1}^{2}}+4\gamma_{21}\left(\tilde{\gamma}_{2}r_{1}-\tilde{\gamma}_{1}r_{2}\frac{\tilde{x}_{2}^{2}}{\tilde{x}_{1}^{2}}\right)\frac{\tilde{x}_{t}^{2}}{x_{t}^{2}}+8\tilde{\gamma}_{t}\gamma_{21}z\frac{\tilde{x}_{2}^{2}}{x_{t}^{2}}\right] / \Delta_{1}
		\end{array}
	\end{equation}
	with
	\begin{equation}\label{e30}
	\begin{array}{lcl}
	\Delta_{1} & =&  8\left(\tilde{\gamma}_{2t}+\frac{\tilde{x}_{2}^2}{\tilde{x}_{1}^{2}}\tilde{\gamma}_{t1}\right)\left(\frac{x_{2}^2}{x_{t}^{2}}\gamma_{1t}a_{2}+\frac{x_{1}^2}{x_{t}^{2}}\gamma_{t2}a_{1}\right)\\& & +2\frac{\tilde{x}_{t}^2}{\tilde{x}_{1}^{2}}\tilde{\gamma}_{21}\left(\frac{x_{2}^2}{x_{t}^{2}}\gamma_{1t}b_{2}+\frac{x_{1}^2}{x_{t}^{2}}\gamma_{t2}b_{1}\right)+8\frac{\tilde{x}_{t}^2}{\tilde{x}_{1}^{2}}\gamma_{12}\tilde{\gamma}_{21}d  \\& & +4\gamma_{12}\left(\tilde{\gamma}_{2t}c_{1}+\frac{\tilde{x}_{2}^2}{\tilde{x}_{1}^{2}}\gamma_{t1}c_{2}\right)
	\end{array}
	\end{equation}
		$\bullet$ For the mode $n=2$ :	
		\begin{equation}\label{e31}
		T_{2}^{11} =-\frac{i x_{1}^{3}}{3}  B_{2}^{11},\quad  T_{2}^{12} =-\frac{i x_{1}^{2}x_{2}}{3}  B_{2}^{12},\quad T_{1}^{1t} =-\frac{i x_{1}^{2}x_{t}}{3}  B_{2}^{1t}
		\end{equation}  
		\begin{equation}\label{e32}
		T_{2}^{21} =-\frac{i  x_{1}x_{2}^{2}}{3}  B_{2}^{21},\quad  T_{2}^{22} =-\frac{i x_{2}^{3}}{3}  B_{2}^{22},\quad T_{2}^{2t} =-\frac{i x_{t}x_{2}^{2}}{3}  B_{2}^{2t}
		\end{equation}
		\begin{equation}\label{e33}
		T_{2}^{t1} =-\frac{i  x_{1}x_{t}^{2}}{3}  B_{2}^{t1},\quad  T_{2}^{t2} =-\frac{i  x_{2}x_{t}^{2}}{3}  B_{2}^{t2},\quad T_{2}^{tt} =-\frac{i x_{t}^{3}}{3}  B_{2}^{tt}
		\end{equation}
		where
		\begin{equation}\label{e34}
		B_{2}^{11} =4\frac{x_{1}^2}{x_{t}^2}\left[3\tilde{\gamma}_{2t}\gamma_{2t}s_{1}+3\tilde{\gamma}_{1t}\gamma_{t2}s_{2}\frac{\tilde{x}_{2}^2}{\tilde{x}_{1}^2}+2\tilde{\gamma}_{12}\gamma_{t2}w\frac{\tilde{x}_{t}^2}{\tilde{x}_{1}^2}\right]/\Delta_{2}
		\end{equation}
		\begin{equation}\label{e35}
		B_{2}^{12} =4\frac{x_{2}^2}{x_{t}^2}\left[3\tilde{\gamma}_{t2}\gamma_{1t}s_{1}+3\tilde{\gamma}_{1t}\gamma_{1t}s_{2}\frac{\tilde{x}_{2}^2}{\tilde{x}_{1}^2}+2\tilde{\gamma}_{12}\gamma_{12}w\frac{\tilde{x}_{t}^2}{\tilde{x}_{1}^2}\right]/\Delta_{2}
		\end{equation}
		\begin{equation}\label{e36}
		B_{2}^{1t} =2\left[3\tilde{\gamma}_{t2}\gamma_{12}s_{1}+3\tilde{\gamma}_{1t}\gamma_{12}s_{2}\frac{\tilde{x}_{2}^2}{\tilde{x}_{1}^2}+2\tilde{\gamma}_{12}\gamma_{12}w\frac{\tilde{x}_{t}^2}{\tilde{x}_{1}^2}\right]/\Delta_{2}
		\end{equation}
		\begin{equation}\label{e37}
		B_{2}^{21} =4\frac{x_{1}^2}{x_{t}^2}\left[3\tilde{\gamma}_{t2}\gamma_{t2}s_{1}+3\tilde{\gamma}_{t1}\gamma_{t2}s_{2}\frac{\tilde{x}_{2}^2}{\tilde{x}_{1}^2}+2\tilde{\gamma}_{12}\gamma_{t2}w\frac{\tilde{x}_{t}^2}{\tilde{x}_{1}^2}\right]/\Delta_{2}
		\end{equation}
		\begin{equation}\label{e38}
		B_{2}^{22} =4\frac{x_{2}^2}{x_{t}^2}\left[3\tilde{\gamma}_{2t}\gamma_{t1}s_{1}+3\tilde{\gamma}_{1t}\gamma_{1t}s_{2}\frac{\tilde{x}_{2}^2}{\tilde{x}_{1}^2}+2\tilde{\gamma}_{12}\gamma_{12}w\frac{\tilde{x}_{t}^2}{\tilde{x}_{1}^2}\right]/\Delta_{2}
		\end{equation}
		\begin{equation}\label{e39}
		B_{2}^{2t} =2\left[3\tilde{\gamma}_{t2}\gamma_{12}s_{1}+3\tilde{\gamma}_{1t}\gamma_{12}s_{2}\frac{\tilde{x}_{2}^2}{\tilde{x}_{1}^2}+2\tilde{\gamma}_{12}\gamma_{12}w\frac{\tilde{x}_{t}^2}{\tilde{x}_{1}^2}\right]/\Delta_{2}
		\end{equation}
		\begin{equation}\label{e40}
		B_{2}^{t1} =12\frac{x_{1}^2}{x_{t}^2}\left[3\tilde{\gamma}_{2t}\gamma_{2t}s_{1}+3\tilde{\gamma}_{1t}\gamma_{t2}s_{2}\frac{\tilde{x}_{2}^2}{\tilde{x}_{1}^2}+2\tilde{\gamma}_{12}\gamma_{t2}w\frac{\tilde{x}_{t}^2}{\tilde{x}_{1}^2}\right]/\Delta_{2}
		\end{equation}
		\begin{equation}\label{e41}
		B_{2}^{t2} =12\frac{x_{2}^2}{x_{t}^2}\left[3\tilde{\gamma}_{t2}\gamma_{1t}s_{1}+3\tilde{\gamma}_{1t}\gamma_{1t}s_{2}\frac{\tilde{x}_{2}^2}{\tilde{x}_{1}^2}+2\tilde{\gamma}_{12}\gamma_{1t}w\frac{\tilde{x}_{t}^2}{\tilde{x}_{1}^2}\right]/\Delta_{2}
		\end{equation}
		\begin{equation}\label{e42}
		B_{2}^{tt} =6\left[3\tilde{\gamma}_{t2}\gamma_{12}s_{1}+3\tilde{\gamma}_{1t}\gamma_{12}s_{2}\frac{\tilde{x}_{2}^2}{\tilde{x}_{1}^2}+2\tilde{\gamma}_{12}\gamma_{12}w\frac{\tilde{x}_{t}^2}{\tilde{x}_{1}^2}\right]/\Delta_{2}
		\end{equation}
		with 
		\begin{equation}\label{e43}
		\begin{array}{lcl}
		\Delta_{2} &= & 12\left(\gamma_{t2}\tilde{\gamma}_{2t}e_{1}\frac{x_{1}^{2}}{x_{t}^2}+\gamma_{t1}\tilde{\gamma}_{1t}e_{2}\frac{x_{2}^{2}}{x_{t}^2}\frac{\tilde{x}_{2}^2}{\tilde{x}_{1}^{2}}\right)\\ & & +12\left(\gamma_{1t}\tilde{\gamma}_{2t}f_{21}\frac{x_{2}^{2}}{x_{t}^2}+\gamma_{t2}\tilde{\gamma}_{t1}f_{12}\frac{x_{1}^{2}}{x_{t}^2}\frac{\tilde{x}_{2}^2}{\tilde{x}_{1}^{2}}\right)+9\gamma_{12}\left(\tilde{\gamma}_{2t}g_{1}+\tilde{\gamma}_{t1}g_{2}\frac{\tilde{x}_{2}^2}{\tilde{x}_{1}^{2}}\right)\\ & & +4\tilde{\gamma}_{12}\left(\gamma_{2t}h_{1}\frac{x_{1}^{2}}{x_{t}^2}+\gamma_{t1}h_{2}\frac{x_{2}^{2}}{x_{t}^2}\right)\frac{\tilde{x}_{t}^2}{\tilde{x}_{1}^{2}}+12\gamma_{21}\tilde{\gamma}_{12}l\frac{\tilde{x}_{t}^2}{\tilde{x}_{1}^{2}}
		\end{array}
		\end{equation}
\section{Effective wavenumber}
Let  $n_{0}$ be the number of spheres per unit volume.  The model of Lupp\'e, Conoir and Norris  formula \cite{16} of the  effective longitudinal wavenumber  $\xi_{\alpha}$ ($\alpha=1,\;2$) for the low concentration ($\frac{n_{0}}{k_{\alpha}^{2}}<< 1$) is :
 \begin{equation}\label{e44}
 	\xi _{\alpha }^{2} =k_{\alpha }^{2} +n_{0} \delta _{1}^{\alpha } +n_{0}^{2} \left(\delta _{2}^{} +\delta _{2}^{\alpha \left(c\right)} \right)+O(n_{0}^{3})
 \end{equation}
 \begin{equation}\label{e45}
 	\delta _{1}^{\alpha } =\frac{4\pi}{i k_{\alpha}}\sum_{n=0}^{+\infty}(2n+1)T_{n}^{\alpha \alpha},
 \end{equation}
 \begin{equation}\label{e46}
 	\delta _{2}^{\alpha } =-\frac{1}{2} \left(\frac{4\pi }{k_{\alpha }^{} } \right)^{4} \sum _{n,m=0}^{+\infty }K_{nm} T_{n}^{\alpha \alpha } T_{m}^{\alpha \alpha }
 \end{equation}
 \begin{equation}\label{e47}
 	\delta _{2}^{\alpha (c)} =-\frac{1}{2}\left(\frac{4\pi}{k_{\alpha}}\right)^{4}\sum_{n,\; m=0}^{+\infty}\sum_{\beta \neq \alpha}\frac{2k_{\alpha}^3}{k_{\beta}\left(k_{\alpha}^{2}-k_{\beta}^{2}\right)}K_{nm}^{\alpha \beta} T_{n}^{\alpha \beta}T_{m}^{\beta\alpha }
 \end{equation} 
 \begin{equation}\label{e48}
 	K_{nm} =\left(\frac{1}{4\pi } \right)^{{3\mathord{\left/ {\vphantom {3 2}} \right. \kern-\nulldelimiterspace} 2} } \sqrt{\left(2n+1\right)\left(2m +1\right)} \sum _{q}q\sqrt{2q+1} G\left(n,0\; ;\; m ,0\; ;\; q\right)
 \end{equation}
 \begin{equation}\label{e49}
 K_{nm}^{\alpha \beta} =\left(\frac{1}{4\pi } \right)^{{3\mathord{\left/ {\vphantom {3 2}} \right. \kern-\nulldelimiterspace} 2} } \sqrt{\left(2n+1\right)\left(2m +1\right)} \sum _{q}\left(\frac{k_{\alpha}}{k_{\beta}}\right)^{q}\sqrt{2q+1} G\left(n,0\; ;\; m ,0\; ;\; q\right)
 \end{equation}
 \begin{equation}\label{e50}
 	G\left(0,n;0,\nu ;q\right)=\frac{G\left(n,0\; ;\; \nu ,0\; ;\; q\right)}{\sqrt{\frac{ \left(2n+1\right)\left(2\nu +1\right) }{ 4\pi \left(2q+1\right)}}}
 \end{equation} 
 where $\beta=1,\;2,\;t$.  $G$ is the Gaunt coefficient \cite{12}. The sum on the index $q$ runs from $|n-m|$ to $n+m$  in steps of two, with $n+m+q$ even. 
$\delta _{1}^{\alpha } $ and $\delta _{2}^{} $ only use scattering coefficients $T_{n}^{\alpha \alpha } $ while $\delta _{2}^{\alpha \left(c\right)} $ involves the mode conversions  scattering coefficients $ T_{n}^{\alpha \beta} $. The values of the first coefficients $K_{nm}$ and $K_{nm}^{\alpha \beta}$ ($K_{mn}=K_{nm}$, $K_{mn}^{\alpha \beta}=K_{nm}^{\alpha \beta}$)   are 
\begin{equation}\label{e51}
	K_{00}=0,\; K_{01}=\frac{3}{16\pi^2},\; K_{11}=\frac{12}{16\pi^2},\; K_{12}=\frac{33}{16\pi^2}  
\end{equation}
\begin{equation}
	K_{02}=\frac{10}{16\pi^2},\; K_{22}=\frac{460}{7\times 16\pi^2}
\end{equation}
\begin{equation}\label{e53}
K_{00}^{\alpha \beta}=\frac{1}{16\pi^2},\, K_{01}^{\alpha\beta}=\frac{3}{16\pi^2}\frac{k_{\alpha}}{k_{\beta}},\, K_{11}^{\alpha \beta}=\frac{3}{16\pi^2}\left(1+2\left(\frac{k_{\alpha}}{k_{\beta}}\right)^2\right)
\end{equation}
\begin{equation}\label{e54}
 K_{02}^{\alpha\beta}=\frac{5}{16\pi^2}\left(\frac{k_{\alpha}}{k_{\beta}}\right)^2,\;K_{12}^{\alpha\beta}=\frac{3}{16\pi^2}\frac{k_{\alpha}}{k_{\beta}}\left(2+3\left(\frac{k_{\alpha}}{k_{\beta}}\right)^2\right)
\end{equation}
\begin{equation}
	 K_{22}^{\alpha \beta}=\frac{1}{ 16\pi^2}\left(5+\frac{50}{7}\left(\frac{k_{\alpha}}{k_{\beta}}\right)^2+\frac{90}{7}\left(\frac{k_{\alpha}}{k_{\beta}}\right)^4\right)
\end{equation}
 Considering the developments at low frequencies of the scattering coefficients, we obtain the following reduced form of equation \eqref{e44}:
   \begin{equation}\label{e55}
  \begin{array}{l} 	\left(\frac{\xi_{\alpha}}{k_{\alpha}}\right)^2\approx 1+\Phi\left[B_{0}^{\alpha \alpha}-1-3B_{1}^{\alpha \alpha}-5 B_{2}^{\alpha \alpha}\right]+\Phi^{2}\left[(1-B_{0}^{\alpha\alpha} )(3 B_{1}^{\alpha\alpha}+10 B_{2}^{\alpha\alpha})\right.\\ \left. \qquad \qquad+6\left(B_{1}^{\alpha\alpha}\right)^2+ 33 B_{1}^{\alpha\alpha}B_{2}^{\alpha\alpha}+\frac{230}{7}\left(B_{2}^{\alpha\alpha}\right)^2\right]\\ \qquad \qquad-\frac{\Phi^2}{2}\left[N^{\alpha\beta}\left[-\left(\frac{k_{\beta}}{k_{\alpha}}\right)^3B_{0}^{\beta\alpha}B_{0}^{\alpha \beta}+3\left(\frac{k_{\beta}}{k_{\alpha}}B_{1}^{\beta\alpha}B_{0}^{\alpha \beta}+\left(\frac{k_{\beta}}{k_{\alpha}}\right)^3 B_{0}^{\beta\alpha}B_{1}^{\alpha \beta} \right)\right. \right. \\ \left. \left.\qquad \qquad -3\left(\frac{k_{\beta}}{k_{\alpha}}\right)^3 \left(1+2\left(\frac{k_{\alpha}}{k_{\beta}}\right)^2\right)B_{1}^{\beta\alpha}B_{1}^{\alpha \beta}\right. \right. \\ \left. \left.\qquad \qquad-\left(6\frac{k_{\beta}}{k_{\alpha}}+9\frac{k_{\alpha}}{k_{\beta}}\right) \left(\left(\frac{k_{\beta}}{k_{\alpha}}\right)^2 B_{2}^{\beta\alpha}B_{1}^{\alpha\beta}+B_{1}^{\beta\alpha}B_{2}^{\alpha\beta}\right)\right. \right.\\ \left. \left.\qquad \qquad -\left(5+\frac{50}{7}\left(\frac{k_{\alpha}}{k_{\beta}}\right)^2+\frac{90}{7}\left(\frac{k_{\alpha}}{k_{\beta}}\right)^4\right)\left(\frac{k_{\beta}}{k_{\alpha}}\right)^3 B_{2}^{\beta\alpha}B_{2}^{\alpha\beta} \right] \right. \\ \left. \qquad \qquad-N^{\alpha t}\left[3\left(1+2\left(\frac{k_{\alpha}}{k_{t}}\right)^2\right)\left(\frac{k_{t}}{k_{\alpha}}\right)^3 B_{1}^{t\alpha}B_{1}^{\alpha t}\right. \right.\\ \left. \left.\qquad \qquad+\left(6\frac{k_{t}}{k_{\alpha}}+9\frac{k_{\alpha}}{k_{t}}\right)\left( B_{2}^{t\alpha} B_{1}^{\alpha t}+ B_{1}^{t\alpha} B_{2}^{\alpha t}\right)\right. \right. \\ \left. \left.\qquad \qquad+\left(5+\frac{50}{7}\left(\frac{k_{\alpha}}{k_{t}}\right)^2+\frac{90}{7}\left(\frac{k_{\alpha}}{k_{t}}\right)^4\right)\left(\frac{k_{t}}{k_{\alpha}}\right)^3  B_{2}^{t\alpha} B_{2}^{\alpha t}\right]\right]\end{array}
  \end{equation}
 where $N^{\alpha \beta}=\frac{2k_{\alpha}^3}{k_{\beta}\left(k_{\alpha}^{2}-k_{\beta}^{2}\right)}$ ($\alpha\neq \beta=1,\;2$) and $N^{\alpha\, t }=\frac{2k_{\alpha}^3}{k_{t}\left(k_{\alpha}^{2}-k_{t}^{2}\right)}$ ($\alpha=1,\;2$). $\Phi=\frac{4\pi n_{0}a^{3}}{3}$ is volume fraction of spheres. At very low frequencies, equation \eqref{e55} will be used later to determine the effective mass density and bulk modulus of the heterogeneous poroelastic medium  by considering coherent fast wave while   the behavior of the  coherent  slow wave at very low frequencies leads only  to  the extraction of the effective diffusion coefficient \cite{22}. 

\newpage
\section{Numerical results}
\subsection{Scattering coefficients}
\begin{figure}[htbp!]
\begin{minipage}[c]{0.5\linewidth}
	\includegraphics[scale=0.4]{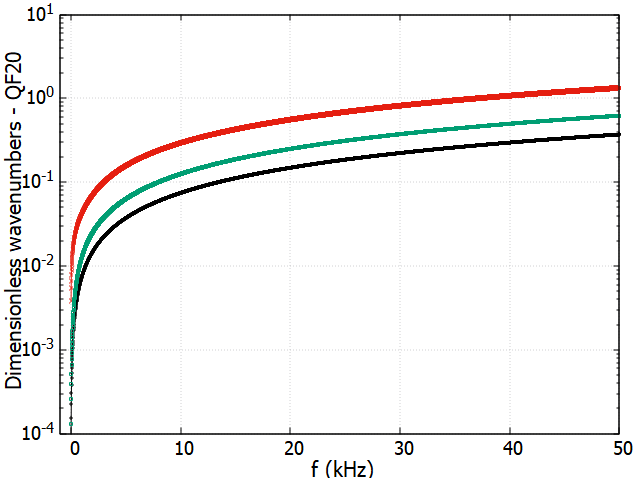}
\end{minipage}
\begin{minipage}[c]{0.1\linewidth}
	\includegraphics[scale=0.4]{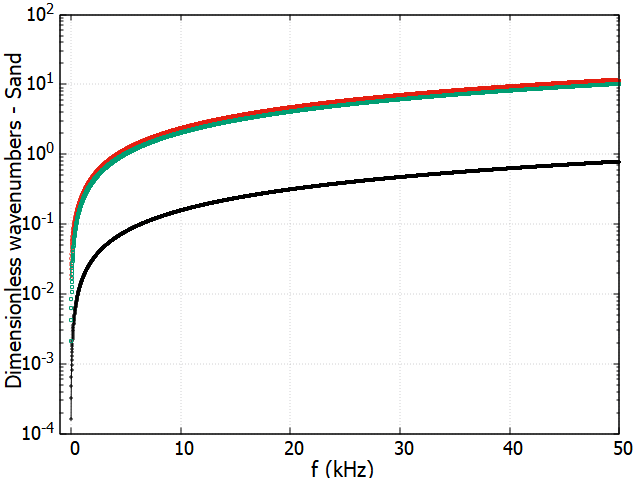}
\end{minipage}
	\caption{Modulus of the dimensionless wavenumbers $x_{\alpha}$  fast wave ($\alpha=1$, black lines), slow wave ($\alpha=2$, red lines)  and shear wave ($\alpha=t$, green lines)}
	\label{f2}
\end{figure}

We consider two porous media QF20 and Stoll sand saturated by the same fluid $\rho$ (table 1). The radius of the spherical obstacles is $a=4\times 10^{-3}$ and the volume fraction of the spheres is $\phi=20\%$. As the scattering coefficients depend on the non dimensional wavenumbers $x_{\alpha}$ ($\alpha=1,\;2,\; t$), we first construct the evolution of the module $|x_{\alpha}|$ as a function of the frequencies in the matrix without  inclusion. Fig.\ref{f2} shows  the case of the matrix QF20 when $|x_{\alpha}| <1$  $(\alpha=1,\;t)$ for  frequency between 0 and 50 kHz. For slow wave we have $|x_{2}|<1$ for frequencies lower than 37 kHz. For the case of the matrix Stoll sand $|x_{1}|$ is less than 1 while $|x_{2}|$ and $|x_{t}|$ are greater than 1 for frequencies greater than 4.38 kHz and 4.89 kHz respectively. The low frequency assumption is violated and this influences  the approximation of the scattering coefficients. However we can admit a construction domain for frequencies between 0 and 50.000 Hz because the develpment in Taylor series gives restricted domains than that allowed by experiments \cite{13}. We recall that smaller sphere radius allows for better approximation and wider construction domains.

\begin{figure}[htbp!]
	\begin{minipage}[c]{0.52\linewidth}
		\includegraphics[scale=0.6]{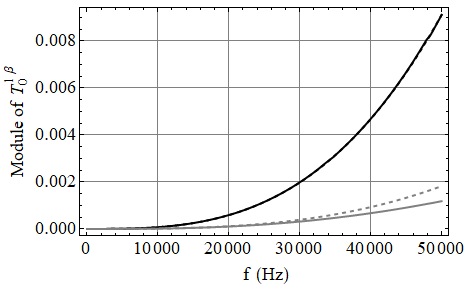}
	\end{minipage}
	\begin{minipage}[c]{0.1\linewidth}
		\includegraphics[scale=0.57]{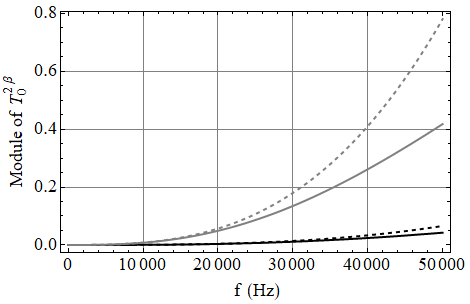}
	\end{minipage}
	\\
		\begin{minipage}[c]{0.52\linewidth}
			\includegraphics[scale=0.57]{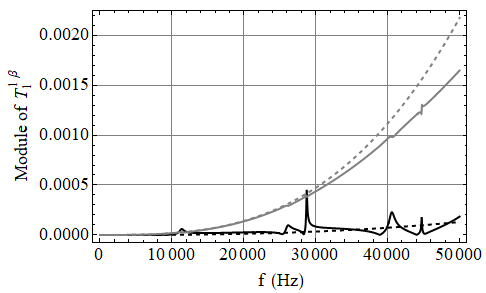}
		\end{minipage}		
		\begin{minipage}[c]{0.1\linewidth}
			\includegraphics[scale=0.57]{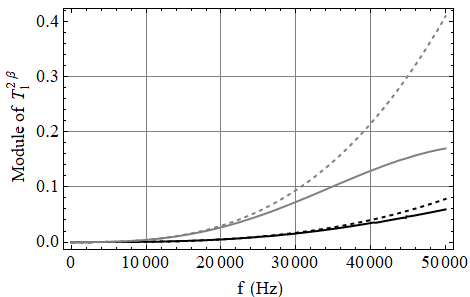}
		\end{minipage}	
		\\
		
		\begin{minipage}[c]{0.52\linewidth}
			\includegraphics[scale=0.56]{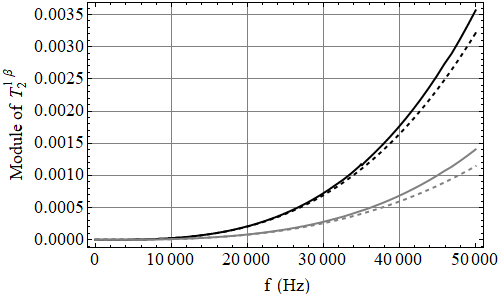}
		\end{minipage}		
		\begin{minipage}[c]{0.1\linewidth}
			\includegraphics[scale=0.57]{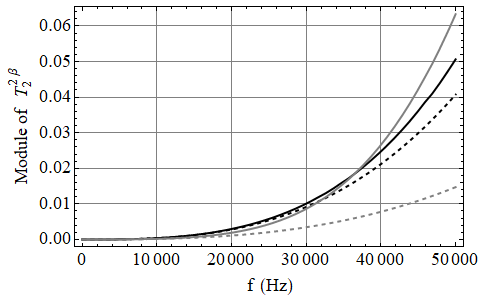}
		\end{minipage}
		\caption{\small{ Module of exact (solid lines) and approximated (dashedlines)  $T_{n}^{1 \beta}$ (left side) and $T_{n}^{2 \beta}$ (right side) plotted versus frequency (Black color $\beta=1$,  Gray color for $\beta= 2$)}.}
		\label{f3}
			
\end{figure}

\noindent  Fig.\ref{f3}  shows the behaviour of  the modulus of $T_{n}^{\alpha \beta}$ $(\alpha,\; \beta=1,\;2,\quad n=0,\,1,\,2 )$ versus the frequency  in the case of sand sphere in QF20 matrix. The solid curves  correspond to the exact scattering coefficients and the  dashed ones to the low frequency approximation. The divergences between the exact and approximate scattering coefficients are greater in the case of the incident slow wave than the case of the incident fast wave. This is due to  fact that $|x_{2}|>1$ as soon as $f>37.000$ Hz and we have the condition $|x_{2}|<<1$ if and only if $f<10.000$ Hz.  The aproximation of $T_{0}^{11}$ is good in the whole  frequency range.

\newpage
\subsection{Effective wavenumbers}
\begin{figure}[htbp!]
	\begin{minipage}[c]{0.5\linewidth}
		\includegraphics[scale=0.4]{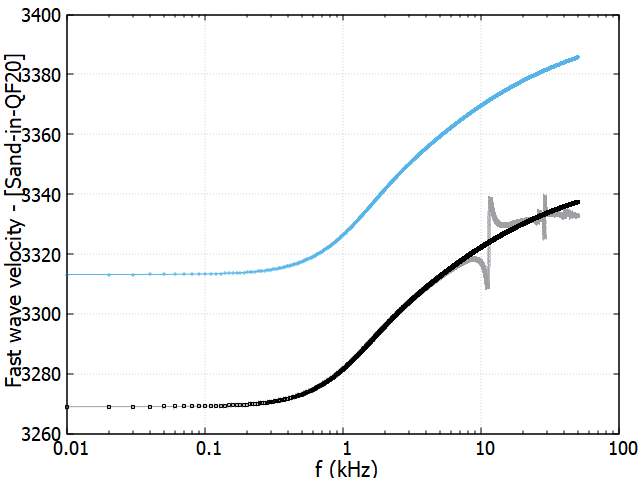}
	\end{minipage}
	\begin{minipage}[c]{0.1\linewidth}
		\includegraphics[scale=0.4]{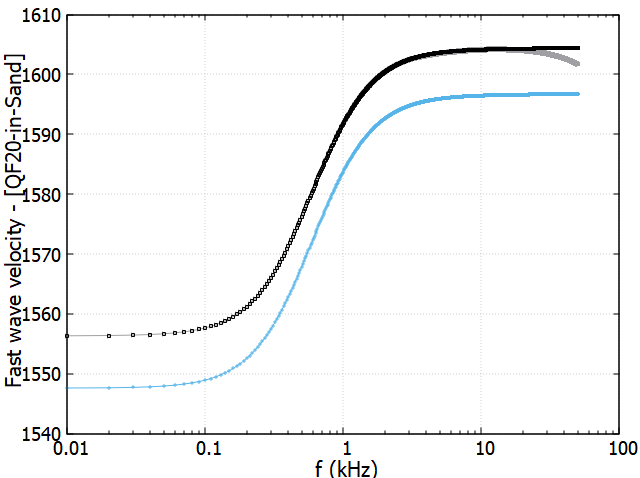}
	\end{minipage}
	\\
	\begin{minipage}[c]{0.5\linewidth}
		\includegraphics[scale=0.4]{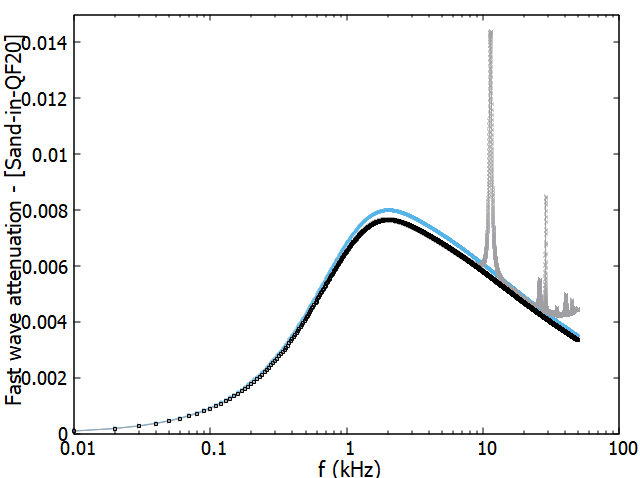}
	\end{minipage}		
	\begin{minipage}[c]{0.1\linewidth}
		\includegraphics[scale=0.4]{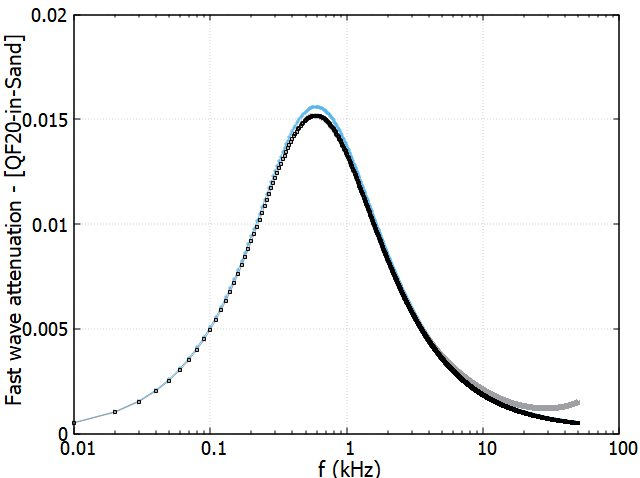}
	\end{minipage}	

\caption{\small{Phase velocity and attenuation of the fast wave  versus frequency (in logarithmic scale) in a heterogeneous porous medium : case of sand spheres in QF20 on the left, and case of QF20 spheres in sand  on the right. volume fraction is $\phi=20\%.$. Solid gray lines
represents the exact effective phase velocity and attenuation, Black lines represents the approximate effective phase velocity and attenuation. The blue curves correspond to poroelastic matrix free of scatterers. } }
\label{f4}
	
\end{figure}

\begin{figure}[htbp!]
	\begin{minipage}[c]{0.5\linewidth}
		\includegraphics[scale=0.4]{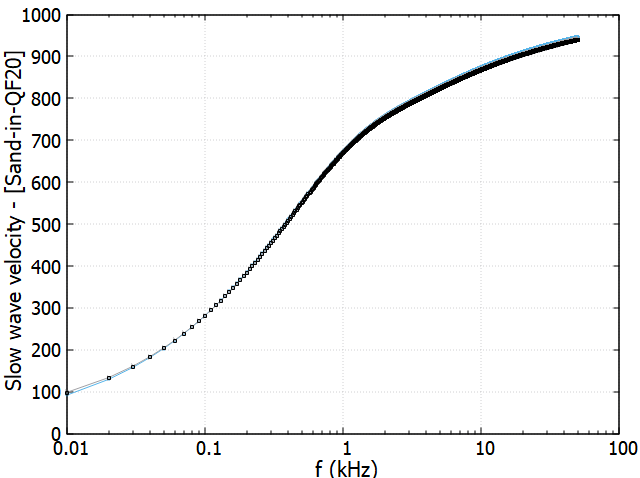}
	\end{minipage}
	\begin{minipage}[c]{0.1\linewidth}
		\includegraphics[scale=0.4]{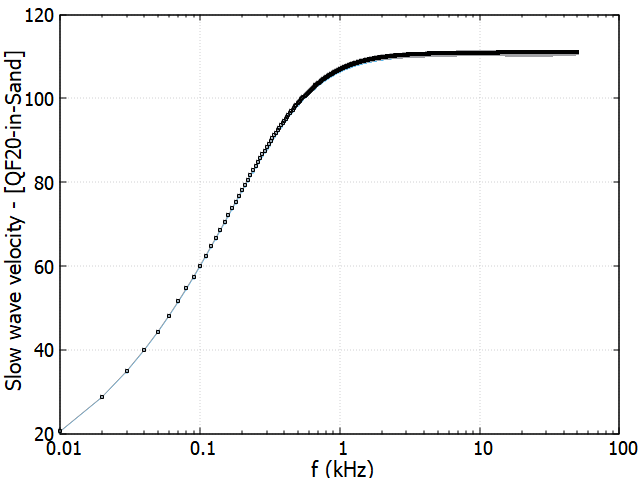}
	\end{minipage}
	\\
	\begin{minipage}[c]{0.5\linewidth}
		\includegraphics[scale=0.4]{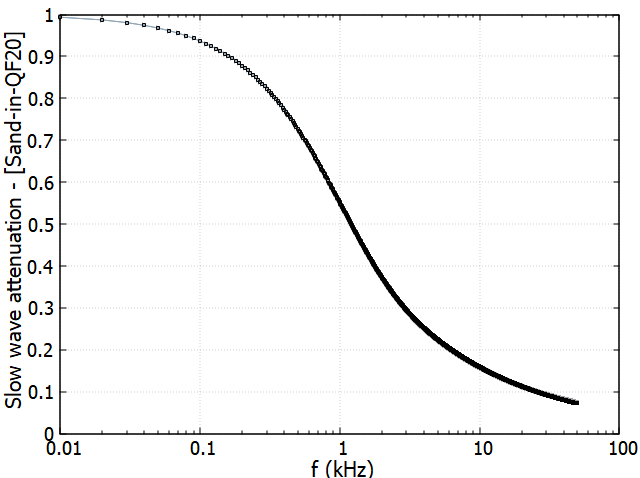}
	\end{minipage}		
	\begin{minipage}[c]{0.1\linewidth}
		\includegraphics[scale=0.4]{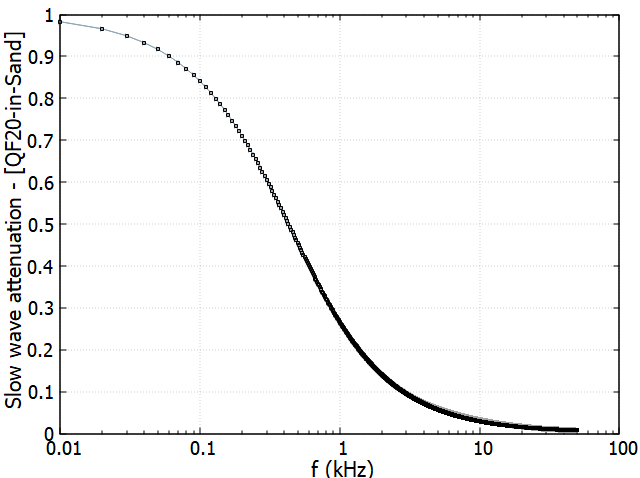}
	\end{minipage}	
	
	\caption{\small{Phase velocity and attenuation of the slow wave  versus frequency (in logarithmic scale) in a heterogeneous porous medium : case of sand spheres in QF20 on the left, and case of QF20 spheres in sand  on the right. volume fraction is $\phi=20\%$. Solid gray lines represents the exact effective phase velocity and attenuation, Black lines represents the approximate effective phase velocity and attenuation. The blue curves correspond to poroelastic matrix free of scatterers. } }
	\label{f5}
	
\end{figure}

Let us consider the following two materials : 
\begin{itemize}
\item a poroelastic QF20 matrix containing a random distibution of \\poroelastic sand spheres refered to as [Sand-in-QF20],
\item  a poroelastic sand matrix containing a random distibution of \\poroelastic QF20 spheres refered to as [QF20-in-Sand].
\end{itemize}
The phase velocity and attenuation for  fast and slow waves  in the absence of scatterers in the random medium are defined by $v_{\alpha}=\Re\left(\frac{\omega}{k _{\alpha} } \right)$ and $\mathcal{A}_{\alpha}=\frac{\Im\left(k _{\alpha} \right)}{\Re\left(k _{\alpha} \right)} $ ($\alpha =1,\; 2$), respectively illustrated with blue solid lines.\\
 The phase velocity and attenuation for fast and slow waves  in the presence of scatterers  in the effective matrix are defined by $v_{\alpha,\, eff}=\Re\left(\frac{\omega}{\xi_{\alpha} } \right)$ and $\mathcal{A}_{\alpha,\, eff}=\frac{ \Im \left(k _{\alpha} \right)}{\Re\left(k _{\alpha} \right)} $
 ($\alpha =1,\; 2$), respectively.  These   two  effective quantities are illustrated by gray solid lines and black solid lines for the exact and approximate formulas of  $\xi_{\alpha}$, respectively.  Fig.\ref{f4} and Fig.\ref{f5} illustrate the variations of the velocities and attenuations  for [Sand-in-QF20] on the left and for [QF20-in-Sand] on the right of the two longitudinal wave. The simulation interval is 0.01 kHz to 50 kHz. Fig.\ref{f4} shows the variation of the fast wave. We observe a very good agreement  of the effective velocities and attenuations below 10 kHz for [sand-in-QF20] or [QF20-in-sand].\\ $v_{1,\, eff}<v_{1}$ for [Sand-in-QF20] and $v_{1,\, eff}>v_{1}$ for [QF20-in-sand]. \\ Attenuation $\mathcal{A}_{1}$  differs little  from effective attenuations $\mathcal{A}_{1,\,eff}$ whatever the case. The maximum  of attenuations is  reached at $f=0,6$ kHz for [QF20-in-sand] while it is reached  at  $f=2,89$ kHz for  [Sand-in-QF20]. In the case of  [Sand-in-QF20], from 10 kHz we observe the oscillations   of the exact effective phase velocity of fast wave due to  the resonance. In the case of [QF20-in-sand] we have no peaks because the  sand matrix is soft matrix. The same phenomenon has been observed by Gnadjro and al. in the case of poroelastic cylinders in the poroelastic matrix\cite{27} (See also Liu and al. \cite{8}). 

Fig.\ref{f5} illustrates the behavior of slow wave. The exact and approximate curves of the velocities and attenuations  are practically the same in the whole frequency range for the two cases.  No oscillation is observed is the case of  slow wave because it  highly diffusive at low
frequency.

\newpage

\section{Effective mass density and bulk modulus of the coherent fast wave}
Neglecting the second order term in $\Phi$, the equation \eqref{e55} is reduced to the form
\begin{equation}\label{e56}
	\left(\frac{\xi_{1}}{k_{1}}\right)^2\approx \left[1+\Phi\left(B_{0}^{11}-1-5 B_{2}^{11}\right)\right]\left[1-3\Phi B_{1}^{11}\right]
	\end{equation}
The goal of this section is to extract from the formula of the equation \eqref{e56} the effective mass density and bulk modulus at the static limit such as the angular frequency $\omega$ tends to zero.  Let be $\omega _{c} =\frac{\eta }{ \rho _{0} \kappa }$ and $\tilde{\omega}_{c} =\frac{\tilde{\eta} }{ \tilde{\rho_{0} }\tilde{\kappa} }$ be the  characteristic frequencies for matrix and the poroelastic sphere  respectively. At the static limite  $\omega$ is very small compared to the mimimum of $\omega_{c }$ and $\tilde{\omega}_{c}$. When $\omega$ tends to zero, the wavenumbers in the matrix and in the sphere   are \cite{9}
\begin{equation}\label{e57}
	k_{1}^{2} \simeq \frac{\omega ^{2}\rho}{H} ,\quad k_{2}^{2} \simeq \frac{i\omega \omega _{c} \rho _{0} H }{\left(HM-C^{2} \right)},\quad k_{t}^{2} \simeq \frac{\omega ^{2} \rho}{ \mu }
\end{equation}
\begin{equation}\label{e58}
	\tilde{k}_{1}^{2} \simeq \frac{\omega ^{2}\tilde{\rho}}{\tilde{H}},\quad \tilde{ k}_{2}^{2} \simeq \frac{i\omega \tilde{\omega} _{c} \rho _{0} \tilde{H} }{\left(\tilde{H}\tilde{M}-\tilde{C}^{2} \right)},\quad \tilde{k}_{t}^{2} \simeq \frac{\omega ^{2} \tilde{\rho}}{ \tilde{\mu} }.
\end{equation}
From the equations \eqref{e57}, \eqref{e58} and taking in consideration the equations \eqref{ap2}-\eqref{ap5}, it follows that 
\begin{equation}\label{e59}
	\gamma _{1} \simeq 0, \; \gamma _{2} \simeq -\frac{H}{C} ,\;  \gamma _{t} \simeq 0,\;   
	\tilde{\gamma} _{1} \simeq 0, \; \tilde{\gamma} _{2} \simeq -\frac{\tilde{H}}{\tilde{C}} ,\;  \tilde{\gamma} _{t} \simeq 0.
\end{equation}
From the previous limits,  the following approximations are deduced at static limit:
\begin{equation}\label{e60}
	B_{0}^{11} \simeq \frac{3 H}{3\tilde{H}-4\tilde{\mu}+4\mu}
\end{equation}
\begin{equation}\label{e61}
	\left(B_{0}^{12},B_{0}^{21},B_{0}^{22}\right) \simeq \left(0,0,0\right)
\end{equation}
\begin{equation}\label{e62}
	\left( B_{1}^{11},B_{1}^{12},B_{1}^{1t}\right) \simeq \left(\frac{\rho-\tilde{\rho}}{3\rho},0,\frac{\rho-\tilde{\rho}}{3\rho}\right)
\end{equation}
\begin{equation}\label{e63}
	\left( B_{1}^{21},B_{1}^{22},B_{1}^{2t}\right) \simeq \left(\frac{\rho-\tilde{\rho}}{3\rho},-\frac{1}{2} ,-\dfrac{\rho+2\tilde{\rho}}{6\rho}\right)
\end{equation}
\begin{equation}\label{e64}
	\left( B_{1}^{t1},B_{1}^{t2},B_{1}^{tt}\right) \simeq \left(2\frac{\rho-\tilde{\rho}}{3\rho},0,2\dfrac{\rho-\tilde{\rho}}{3\rho}\right)
\end{equation}
\begin{equation}\label{e65}
	\left( B_{2}^{11},B_{2}^{12},B_{2}^{1t}\right) \simeq \left(-\frac{4\mu \mathcal{B}_{1}}{\mathcal{B}_{2}},0,-\frac{2H \mathcal{B}_{1}}{\mathcal{B}_{2}}\right)
\end{equation}
\begin{equation}\label{e67}
	\left( B_{2}^{21},B_{2}^{22},B_{2}^{2t}\right) \simeq \left(-\frac{4\mathcal{B}_{1}}{\mathcal{B}_{2}},0,-\frac{4H \mathcal{B}_{1}}{\mathcal{B}_{2}}\right)
\end{equation}
\begin{equation}\label{e68}
	\left( B_{2}^{t1},B_{2}^{t2},B_{2}^{tt}\right) \simeq \left(\frac{4\mathcal{B}_{1}}{\mathcal{B}_{2}},0,\frac{4H \mathcal{B}_{1}}{\mathcal{B}_{2}}\right)
\end{equation}
where 
\begin{equation}
	\mathcal{B}_{1}=3\tilde{\mu}\left(7(\tilde{H}-2\tilde{\mu})(\mu-\tilde{\mu})+8\mu^2-2\mu\tilde{\mu}-6\tilde{\mu}^2\right)+2\tilde{H}(8\mu^2-9\mu\tilde{\mu}+\tilde{\mu}^2)
\end{equation}
\begin{equation}
 \begin{array}{lcl} 
	\mathcal{B}_{2}= 12\mu \tilde{\mu}(\mu-\tilde{\mu})(12H-7\tilde{H}-8\mu+8\tilde{\mu})+12H\tilde{H}(\mu-\tilde{\mu})(4\mu+\tilde{\mu})\\\qquad+9H\tilde{\mu}\left[7(\tilde{H}-2\tilde{\mu})(3\mu+2\tilde{\mu})+8\mu^2+50\mu\tilde{\mu}+12\tilde{\mu}^{2}\right]\\ \qquad  +4\mu\tilde{H}\left[12(H-2\mu)(2\mu+3\tilde{\mu})+32\mu^2+75\mu\tilde{\mu}-2\tilde{\mu}^{2}\right]
	\end{array}
\end{equation}
Considering the equation \eqref{e57}-\eqref{e58},  we suppose that at the static limit, the  effective wavenumbers formula of fast   wave \cite{18}  is  
$	\xi_{1}^{2}=\frac{\omega^{2} \rho_{1,\,eff}}{H_{1,\,eff}}$ where $\rho_{1,\,eff}$ and $H_{1,\,eff}$  denote the  effective mass density and modulus bulk  of fast coherent waves, respectively. It follows that $\frac{\xi_{1}^{2}}{k_{1}^{2}}=\frac{\rho_{1,\,eff}}{\rho}\times \frac{H}{H_{1,\,eff }}$. By comparing to equation \eqref{e56}, we deduce the effective mass density and bulk modulus
\begin{equation} \label{e69} 
\frac{\rho _{1,\,eff} }{\rho _{1} } =1-\Phi\left(\frac{\rho-\tilde{\rho}}{\rho}\right)
\end{equation}
\begin{equation} \label{e70} 
	\frac{1}{H_{1,\,eff}}=\frac{1-\Phi}{H}+\Phi\left(\frac{3}{3\tilde{H}-4\tilde{\mu}+4\mu}+20\frac{\mu}{H}\frac{\mathcal{B}_{1}}{\mathcal{B}_{2}}\right)
\end{equation} 

\subsection{Poroelastic matrix containing elastic spheres}
In this case, the porosity of the poroelastic sphere becomes low ($\tilde{\phi} \to 0$), then  $\tilde{\rho} \to \tilde{\rho}_{s}$, $\tilde{\mu}\to \tilde{\mu}_{s}$,  $\tilde{H}\to \tilde{\lambda}_{s}+2\tilde{\mu}_{s}$. $\tilde{\mu}_{s}$ and $\tilde{\mu}_{s}$ are Lam\'e constants, $\tilde{\rho}_{s}$ is the mass density. Equation \eqref{e69} to equation \eqref{e70} become
\begin{equation} \label{e71} 
\frac{\rho _{1,\,eff} }{\rho _{1} } =1-\Phi\left(\frac{\rho-\tilde{\rho}_{s}}{\rho}\right)
\end{equation}
\begin{equation} \label{e72} 
\frac{1}{H_{1,\,eff}}=\frac{1-\Phi}{H}+\Phi\left(\frac{3}{3\tilde{\lambda}_{s}+2\tilde{\mu}_{s}+4\mu}+20\frac{\mu}{H}\frac{\mathcal{B}_{1}^{s}}{\mathcal{B}_{2}^{s}}\right)
\end{equation} 
where 
\begin{equation} \label{e73} 
\mathcal{B}_{1}^{s}=\tilde{\lambda}(\mu-\tilde{\mu})(16\mu-19\tilde{\mu})+14\tilde{\mu}(\tilde{\mu}-\mu)(4\mu-\tilde{\mu})
\end{equation} 
\begin{equation} \label{e74} 
\begin{array}{lcl} 
\mathcal{B}_{2}^{s}= 12\mu \tilde{\mu}_{s}(\mu-\tilde{\mu}_{s})(12H-7\tilde{\lambda}_{s}-8\mu-6\tilde{\mu}_{s})\\\qquad+12H(\tilde{\lambda}_{s}+2\tilde{\mu}_{s})(\mu-\tilde{\mu}_{s})(4\mu+\tilde{\mu}_{s})\\\qquad+9H\tilde{\mu}_{s}\left[7\tilde{\lambda}_{s}(3\mu+2\tilde{\mu}_{s})+8\mu^2+50\mu\tilde{\mu}_{s}+12\tilde{\mu}_{s}^{2}\right]\\ \qquad  +4\mu(\tilde{\lambda}_{s}+2\tilde{\mu}_{s})\left[12H(2\mu+3\tilde{\mu}_{s})-16\mu^2+3\mu\tilde{\mu}_{s}-2\tilde{\mu}_{s}^{2}\right]
\end{array}
\end{equation} 
Equations \eqref{e71} and \eqref{e72} are the effective mass density and bulk modulus  in the case of elastic sphere  in a poroelastic matrix given by Kuagbenu and all \cite{22}. We have completed the expression to the term of order 1 in $\Phi$ by $20\frac{\mu}{H}\frac{\mathcal{B}_{1}^{s}}{\mathcal{B}_{2}^{s}}$.
 
\subsection{Perforated poroelastic matrix}
Consider the case of poroelastic matrix containing spherical cavities randomly distributed. When the porosity tends towards zero, we have $\tilde{\phi}\to 1 $, $\tilde{\mu}\to 0$, $\tilde{\rho}\to \tilde{\rho}_{0}$ (mass density of fluid sphere) and $\tilde{H}\to \tilde{K}_{0}$ (bulk modulus of fluid sphere). It  follows that
\begin{equation} \label{e75} 
\frac{\rho _{1,\,eff} }{\rho _{1} } =1-\Phi\left(\frac{\rho-\tilde{\rho}_{0}}{\rho}\right)
\end{equation}
\begin{equation} \label{e76} 
\frac{1}{H_{1,\,eff}}=\frac{1-\Phi}{H}+\Phi\left(\frac{3}{3\tilde{K}_{0}+4\mu}+\frac{20\mu}{H\left(9H-4\mu\right)}\right)
\end{equation}

\subsection{Poroelastic spheres immersed in fluid}
 Poroelastic matrix with  a high porosity ($\phi \to 1$) behaves as fluid medium described by mass density $\rho_{0}$ and bulk modulus $K_{0}$. Then  we obtain  the limits $\mu \to 0$,  $\rho \to \rho_{0}$  and $H\to K_{0}$.
 We deduce the  following effective properties
\begin{equation} \label{e77} 
\frac{\rho _{1,\,eff} }{\rho _{1} } =1-\Phi\left(\frac{\rho_{0}-\tilde{\rho}}{\rho_{0}}\right)
\end{equation}
\begin{equation} \label{e78} 
\frac{1}{H_{1,\,eff}}=\frac{1-\Phi}{H}+\Phi\left(\frac{3}{3\tilde{H}-4\tilde{\mu}}\right)
\end{equation}
Equation \eqref{e78} agrees with equation  (33.s) of Gnadjro and all \cite{25} given for poroelastic spheres immersed in fluid in the case of polydisperse assembly of randomly distributed spheres. Equation \eqref{e77} is not conform to the equation  (32.s) of \cite{25}   because we use the sealed pore conditions.

\section{Conclusion}
In this paper, the propagation of fast and slow wave through a random distribution of identical poroelastics spheres in a poroelastic matrix using Biot's theory is considered.  Using  LCN's model and low frequency limit, explicit expressions for the effective mass density  and bulk modulus for fast coherent wave have been deduced. The effective properties of the material in the  static limit are also derived. The behavior of the slow wave allowed to extract the effective diffusion coefficient  which is not calculated here. The numerical results  of velocities and attenuations showed the influence of  the different porosities of the spherical inclusions and matrix on the longitudinal wave propagation.  This work can be used in non-destructive material testing,  mechanical properties of composite materials.

\section*{Appendix}
\addcontentsline{toc}{section}{Appendix}

\appendix  

\section{Matrix equation}\label{ap1}
The  elements of  matrix equation for a poroelastic sphere in a poroelastic matrix, are given by :

\begin{eqnarray}
m_{n,\,11}&=& x_{1}h_{n}'(x_{1})\\
m_{n,\,12}&=&x_{2}h_{n}'(x_{2})\\
m_{n,\,13}&=&n(n+1) h_{n}(x_{t})\\
m_{n,\,14}&=&-\tilde{x}_{1}j_{n}'(\tilde{x}_{1})\\
m_{n,\,15}&=&-\tilde{x}_{2}j_{n}'(\tilde{x}_{2})\\
m_{n,\,16}&=&-n(n+1) j_{n}(\tilde{x}_{t})
\end{eqnarray}
\begin{eqnarray}
m_{n,\,21}&=& j_{n}(x_{1})\\
m_{n,\,22}&=&j_{n}(x_{2})\\
m_{n,\,23}&=&h_{n}(x_{t})+x_{t} h_{n}'(x_{t})\\
m_{n,\,24}&=& -j_{n}(\tilde{x}_{1})\\
m_{n,\,25}&=&- j_{n}(\tilde{x}_{2})\\
m_{n,\,26}&=&-j_{n}(\tilde{x}_{t})-\tilde{x}_{t} j_{n}'(\tilde{x}_{t})
\end{eqnarray}
\begin{eqnarray}
m_{n,\,31}&=& \gamma_{1} x_{1} h_{n}'(x_{1})\\
m_{n,\,32}&=& \gamma_{2} x_{2} h_{n}'(x_{2})\\
m_{n,\,33}&=& n (n+1)\gamma_{t} h_{n}(x_{t})\\
m_{n,\,34}&=& 0\\
m_{n,\,35}&=& 0\\
m_{n,\,36}&=&0
\end{eqnarray}
\begin{eqnarray}
m_{n,\,41}&=& 0\\
m_{n,\,42}&=& 0\\
m_{n,\,43}&=& 0 \\
m_{n,\,44}&=& \tilde{\gamma}_{1} \tilde{x}_{1} j_{n}'(\tilde{x}_{1})\\
m_{n,\,45}&=& \tilde{\gamma}_{2} \tilde{x}_{1} j_{n}'(\tilde{x}_{2})\\
m_{n,\,46}&=& n(n+1) \tilde{\gamma}_{t} j_{n}(\tilde{x}_{t})
\end{eqnarray}
\begin{eqnarray}
m_{n,\,51}&=& \left(2\mu\left((n+1)n-x_{1}^{2}\right)-x_{1}^{2}\mathcal{H}_{1}\right)h_{n}(x_{1})-4\mu x_{1}h_{n}'(x_{1})\\
m_{n,\,52}&=& \left(2\mu\left((n+1)n-x_{2}^{2}\right)-x_{2}^{2}\mathcal{H}_{2}\right)h_{n}(x_{2})-4\mu x_{1}h_{n}'(x_{2})\\
m_{n,\,53}&=& 2\mu n (n+1)\left( x_{t}h_{n}'(x_{t})-h_{n}(x_{t}) \right)\\
m_{n,\,54}&=& \left(-2\tilde{\mu}\left(n(n+1)-\tilde{x}_{1}^2\right)+\tilde{x}_{1}^2 \tilde{\mathcal{H}}_{1}\right)j_{n}(\tilde{x}_{1}))+4\tilde{\mu}\tilde{x}_{1}j_{n}'(\tilde{x}_{1})\\
m_{n,\,55}&=&\left(-2\tilde{\mu}\left(n(n+1)-\tilde{x}_{2}^2\right)+\tilde{x}_{2}^2 \tilde{\mathcal{H}}_{2}\right)j_{n}(\tilde{x}_{2}))+4\tilde{\mu}\tilde{x}_{2}j_{n}'(\tilde{x}_{2})\\
m_{n,\,56}&=& -2\tilde{\mu}n (n+1)\left( \tilde{x}_{t} j_{n}'(\tilde{x}_{t})-j_{n}(\tilde{x}_{t})\right)
\end{eqnarray}
\begin{eqnarray}
m_{n,\,61}&=& 2 \mu  \left(-h_{n}(x_{1})+x_{1}h_{n}'(x_{1})\right)\\
m_{n,\,62}&=& 2 \mu  \left(-h_{n}(x_{2})+x_{2}h_{n}'(x_{2})\right)\\
m_{n,\,63}&=& 2\mu \left(\left(n(n+1)-x_{t}^2/2-1\right)h_{n}(x_{t})-x_{t}h_{n}'(x_{t})\right) \\
m_{n,\,64}&=& 2 \tilde{\mu}  \left(j_{n}(\tilde{x}_{1})-\tilde{x}_{1}j_{n}'(\tilde{x}_{1})\right)\\
m_{n,\,65}&=& 2 \tilde{\mu}  \left(j_{n}(\tilde{x}_{2})-\tilde{x}_{2}j_{n}'(\tilde{x}_{2})\right)\\
m_{n,\,66}&=& -2\tilde{\mu}\left(\left(n(n+1)-\tilde{x}_{t}^2/2-1\right)j_{n}(\tilde{x}_{t})-\tilde{x}_{t}j_{n}'(\tilde{x}_{t})\right)
\end{eqnarray}

\noindent  The  components of vector $\vec{S_{n}^{\alpha}}$  in \eqref{e27} for an incident longitudinal wave \\ $\alpha=1,\,2$ are  
\begin{equation}
\left\lbrace\begin{array}{l}
m_{n,\,1}= -2 \mu  \left(-j_{n}(x_{\alpha})+x_{1}j_{n}'(x_{\alpha})\right)\\
m_{n,\,2}= -j_{n}(x_{\alpha})\\
m_{n,\,3}= -\gamma_{\alpha} x_{\alpha} h_{n}'(x_{1})\\
m_{n,\,4}= 0\\
m_{n,\,5}= -\left(2\mu\left((n+1)n-x_{\alpha}^{2}\right)-x_{\alpha}^{2}\mathcal{H}_{\alpha}\right)j_{n}(x_{\alpha})+4\mu x_{1}j_{n}'(x_{\alpha})\\
m_{n,\,6}=- 2 \mu  \left(-j_{n}(x_{\alpha})+x_{\alpha}j_{n}'(x_{\alpha})\right)
\end{array}
\right.
\end{equation}
$j_{n}$ is the  spherical Bessel function of the first kind and $h_{n}$ is the spherical Hankel function of the first kind \cite{21}.\\ The coefficients $\gamma_{\alpha}$ and $\tilde{\gamma}_{\alpha}$ are defined :\\
in  the poroelastic matrix 
\begin{equation}\label{ap2}
\gamma_{\alpha}=\frac{H k_{\alpha}^{2}-\rho\omega^{2}}{\rho_{0}\omega^{2}-C k_{\alpha}^{2}}, \quad(\alpha=1,\,2)
\end{equation}
\begin{equation}\label{ap3}
\gamma_{t}=\frac{\mu k_{t}^{2}-\rho\omega^{2}}{\rho_{0}\omega^{2}}
\end{equation}
and in poroelastic sphere
\begin{equation}\label{ap4}
\tilde{\gamma}_{\alpha}=\frac{\tilde{H}\tilde{k}_{\alpha}^{2}-\tilde{\rho}\omega^{2}}{\tilde{\rho_{0}}\omega^{2}-\tilde{C}\tilde{k}_{\alpha}^{2}}, \quad(\alpha=1,\,2)
\end{equation}
\begin{equation}\label{ap5}
\tilde{\gamma}_{t}=\frac{\tilde{\mu}\tilde{k}_{t}^{2}-\tilde{\rho}\omega^{2}}{\tilde{\rho}_{0}\omega^{2}}
\end{equation}
  
\newpage
\bibliographystyle{plain}

\end{document}